\documentclass[submission,copyright,creativecommons]{eptcs}
\providecommand{\event}{PROLE 2014}
\usepackage{breakurl}             

\usepackage{amsfonts}
\usepackage{amssymb}
\usepackage{amsmath}
\usepackage{stmaryrd}
\usepackage{yhmath}
\usepackage{proof}
\usepackage{graphics}
\usepackage[utf8]{inputenc}
\usepackage{hyperref}
\usepackage{url}
\usepackage{ifthen} 

\usepackage[svgnames]{xcolor}

\newcommand{\todoFlag}{ON} 
\newcommand{\dudaFlag}{ON} %
\newcommand{\coment}[1]{{}}

\newcommand{\den}[1]{[\![#1]\!]}
\newcommand{\denn}[1]{[\![\![#1]\!]\!]}
\newcommand{\nc}{\newcommand}

\nc{\ordSus}{\lesssim} 
\nc{\ordPos}{\lesssim} 
\nc{\ordap}{\sqsubseteq} 
\nc{\rw}{\to} 
\nc{\tor}{\to} 
\nc{\crwlto}{\rightarrowtriangle}  
\nc{\clto}{\crwlto}                
\nc{\conscrwl}{\vdash_{\it CRWL}}                
\nc{\clp}{{\cal P} \vdash_{CRWL^{+}}} 
\nc{\cldt}{{\cal P} \vdash_{CRWL^{d}}} 
\nc{\dend}[1]{\den{#1}^d} 
\nc{\dg}[1]{\den{#1}_{CRWL}}
\nc{\cl}{{\cal P} \vdash_{CRWL}}
\nc{\clho}{{\cal P} \vdash_{HOCRWL}}
\nc{\gl}{{\cal P} \vdash_{CRWL_{let}}}
\nc{\dgl}[1]{\den{#1}_{CRWL_{let}}}
\nc{\dcl}[1]{\dgl{#1}}
\nc{\ddgl}[1]{\denn{#1}_{CRWL_{let}}}
\nc{\ddcl}[1]{\ddgl{#1}}
\nc{\f}{{\to^{l}}}    
\nc{\fnf}{\to_{lnf}}    
\nc{\fd}{\to_{lw}}    
\nc{\fnt}{{\to^{L}}} 
\nc{\nr}{\leadsto} 
\nc{\fnr}{{\leadsto^L}}  
\nc{\fnre}{\leadsto^{L^*}}  
\nc{\fnrc}[1]{\leadsto^{L^{#1}}}  
\nc{\fnrl}{{\leadsto^l}}  

\nc{\var}{{\cal V}}
\nc{\ra}{\tor}
\nc{\leqhyp}{\Subset}
\nc{\tot}[1]{{#1}^\tau}
\nc{\tlr}[1]{\widehat{#1}} 
\nc{\jn}{\Join} 
\nc{\tr}{\underline{\mbox{\textbf{t}}}}
\nc{\slt}{\hookleftarrow} 
\nc{\clt}{\hookrightarrow} 
\nc{\con}{{\cal C}}  
\nc{\cnn}[1]{\con[#1]}  
\nc{\cnnp}[1]{\con'[#1]}  
\nc{\nat}{{\mathbb N}}
\nc{\prog}{{\cal P}} 
\nc{\progh}{\hat{\prog}} 
\nc{\progm}{{\cal M}} 
\nc{\progr}{{\cal R}} 
\nc{\ps}{\vdash}    
\nc{\pps}{\vdash_{\pi CRWL}}    
\nc{\pcrwl}{{\it $\pi$CRWL}} 
\nc{\crwl}{{\it CRWL}}
\nc{\denp}[1]{\den{#1}^{pl}} 
\nc{\dens}[1]{\den{#1}^{sg}} 
\nc{\denr}[1]{\den{#1}^{rw}} 
\nc{\pst}[1]{\mi{pST}(#1)} 
\nc{\icsus}{\mi{CSubst}_\perp^?}
\nc{\icterm}{\mi{CTerm}^?} 
\nc{\ppop}{\rightarrowtail} 
\nc{\pordap}{\ordap_{\pi}} 
\nc{\dsord}{\unlhd} 
\nc{\cltop}{\clto} 
\nc{\dsordp}{\dsord_{M}} 
\nc{\pl}{\pi } 
\nc{\cjto}{{\cal A}} 
\nc{\cjtoD}{{\cal S}} 
\nc{\vran}{vran}
\nc{\pos}[1]{{\cal O}(#1)} 
\nc{\posV}[1]{\tilde{{\cal O}}(#1)} 
\nc{\denrw}[1]{\den{#1}_{rw}} 
\nc{\sexp}{\mi{SExp}}
\nc{\esexp}{{\mi{ESExp}}}
\nc{\scterm}{\mi{SCTerm}}
\nc{\escterm}{\mi{ESCTerm}}
\nc{\scsubst}{\mi{SCSubst}}
\nc{\sexpb}{{\sexp}} 
\nc{\trsexpb}{tr\sexpb}
\nc{\sctermb}{{\scterm}} 
\nc{\esctermb}{{\escterm}} 
\nc{\uscterm}[1]{{st}_{#1}}
\nc{\usexp}[1]{{se}_{#1}}
\nc{\ssubst}{\mi{SSubst}}
\nc{\sbt}{} 
\nc{\sls}{\langle} 
\nc{\srs}{\rangle} 
\nc{\se}[1]{\sls {#1} \srs}
\nc{\rcrwl}{CRWL^{rt}} 
\nc{\cltor}{\clto} 
\nc{\etose}[1]{\widetilde{{#1}}} 
\nc{\wetose}[1]{\widetilde{{#1}}} 
\nc{\setoe}[1]{\wideparen{{#1}}} 
\nc{\wsetoe}[1]{\wideparen{{#1}}} 
\nc{\pru}{\Delta} 
\nc{\prem}[1]{\Pi^{{#1}}} 
\nc{\crule}[1]{\textsc{#1}} 
\nc{\cinfer}[3]{\infer[\crule{{#1}}]{{#2}}{{#3}}}
\nc{\mrel}[2]{{#1} \lessdot {#2}} 
\nc{\nmrel}[2]{{#1} \not\lessdot {#2}} 
\nc{\trs}{{TRS}} 
\nc{\ctrs}{CS} 
\nc{\trss}{\trs's} 
\nc{\ctrss}{\ctrs's} 
\nc{\cs}{\mi{CS}} 
\nc{\fs}{FS} 
\nc{\vextra}[1]{\mi{vExtra}(#1)} 
\nc{\vE}{{{\cal V}_e}} 
\nc{\enr}{{\nr^{\pl}}} 
\nc{\cupcon}{\twoheadleftarrow} 
\nc{\scon}{s\con} 
\nc{\dseq}{\lhd | \rhd} 
\nc{\cups}{\cup_s }
\nc{\refp}[1]{\ref{#1} (page \pageref{#1})}
\nc{\gt}[1]{\hat{#1}} 
\nc{\wgt}[1]{\widehat{#1}} 
\nc{\progg}{{\cal G}} 
\nc{\progq}{{\cal Q}} 
\nc{\eqg}{\stackrel{\text{\tiny ?}}{=}} 
\nc{\tosu}{\Rightarrow} 
\nc{\sufail}{{\it fail}} 
\nc{\soladd}{\oplus} 
\nc{\snarr}{\leadsto} 
\nc{\assus}[1]{{\sigma_{#1}}} 
\nc{\osus}[1]{{#1}^{\text{\tiny o}}} 
\nc{\sualg}{{\cal S}} 
\nc{\ewr}{{\ }^*\!\!\leftarrow} 
\nc{\sordSus}{\mathring{\lesssim}} 
\nc{\seqSus}{\mathring{\cong}} 

 \nc{\longtxt}[1]{{#1}} 
 \nc{\shorttxt}[1]{} 
\nc{\citaDemos}{\cite{generatorsProofs} } 

\nc{\mi}[1]{\mathit{#1}}
\nc{\gen}{\mi{gen}}
\nc{\CS}{\mi{CS}}
\nc{\FS}{\mi{FS}}
\nc{\Exp}{\mi{Exp}}
\nc{\CTerm}{\mi{CTerm}}
\nc{\Cntxt}{\mi{Cntxt}}
\nc{\Subst}{\mi{Subst}}
\nc{\coin}{\mi{coin}}
\nc{\dom}{\mi{dom}}
\nc{\est}{\mi{est}}
\nc{\miflat}{\mi{flat}}
\nc{\ese}{\mi{ese}}
\nc{\dup}{\mi{dup}}

\nc{\vs}{}

\nc{\codesize}{\small}

\nc{\todoLong}[1]{\ifthenelse{\equal{\todoFlag}{ON}}{~\\{\fcolorbox{red}{yellow}{\begin{minipage}{.985\textwidth}{#1}\end{minipage}}}}{}
} 
\nc{\todo}[1]{\ifthenelse{\equal{\todoFlag}{ON}}{~\\\textcolor[rgb]{1.00,0.00,0.00}{{\fbox{\begin{minipage}{.985\textwidth}{#1}\end{minipage}}}}}{}
}
\nc{\todobis}[2]{\ifthenelse{\equal{\todoFlag}{ON}}{~\\{\fcolorbox{red}{#1}{\begin{minipage}{.985\textwidth}{#2}\end{minipage}}}}{}
} 
\nc{\duda}[1]{\ifthenelse{\equal{\dudaFlag}{ON}}{~\\{\fcolorbox{yellow}{Azure}{\begin{minipage}{.985\textwidth}{#1}\end{minipage}}}}{}
}

\newtheorem{example}{Example} 
\newtheorem{theorem}{Theorem} 
\newtheorem{lemma}{Lemma} 
\newtheorem{proposition}{Proposition} 
\newtheorem{definition}{Definition} 

\def\squareforqed{\hbox{\rlap{$\sqcap$}$\sqcup$}}
\def\qed{\ifmmode\squareforqed\else{\unskip\nobreak\hfil
\penalty50\hskip1em\null\nobreak\hfil\squareforqed
\parfillskip=0pt\finalhyphendemerits=0\endgraf}\fi}

\title{Lifting Term Rewriting Derivations in Constructor Systems by Using Generators\thanks{Research
supported by MICINN Spanish project \emph{StrongSoft} (TIN2012-39391-C04-04).}}
\author{Adri\'an Riesco
\institute{Departamento de Sistemas Inform\'aticos y Computaci\'on\\
Universidad Complutense de Madrid, Spain}
\email{ariesco@fdi.ucm.es}
\and
Juan Rodr\'iguez-Hortal\'a
\institute{Lambdoop Solutions}
\email{juan.rodriguez@lambdoop.com}
}
\def\titlerunning{Lifting Term Rewriting Derivations by Using Generators}
\def\authorrunning{A. Riesco \& J. Rodr\'iguez-Hortal\'a}

\begin{document}
\maketitle

\begin{abstract}
Narrowing \longtxt{is a procedure that} was first studied in the context of equational E-unification and \longtxt{that} has been used in a wide range of applications.
The classic completeness result due to Hullot states that any term rewriting derivation starting from an instance of an expression can be `lifted' to a narrowing derivation, whenever the substitution employed is normalized. 
In this paper we adapt the generator-based extra-variables-elimination transformation used in functional-logic programming to overcome that limitation, so we are able to lift term rewriting derivations starting from arbitrary instances of expressions. The proposed technique is limited to left-linear constructor systems and to derivations reaching a ground expression. 
We also present a Maude-based implementation of the technique, using natural rewriting for the on-demand evaluation strategy.
\end{abstract}

\maketitle

\section{Introduction}\label{sect:intro}

Narrowing \cite{baader-nipkow} is a procedure that was first studied in the context of equational E-unification and that has been used in a wide range of applications 
\cite{MeseguerT07Reach,GHLR99}. 
Narrowing can be described as a modification of term rewriting in which matching is replaced by unification so, in a derivation starting from a goal expression, 
it is able to deduce the instantiation of the variables of the goal expression that is needed for the computation to progress. 
The key result for the completeness of narrowing w.r.t.\ term rewriting is \emph{Hullot's lifting lemma} \cite{Hullot80}, which states that any term rewriting derivation $e_1\theta \rw^* e_2$  can be \emph{lifted} into 
a narrowing derivation $e_1 \nr^*_\sigma e_3$ such that $e_3$ and $\sigma$ are more general than $e_2$ and $\theta$---w.r.t.\ to the usual instantiation preorder~\cite{BaaderS01UnifTheory}, and for the variables involved in the derivations---, provided that the starting substitution $\theta$ is normalized~\cite{DBLP:journals/aaecc/MiddeldorpH94}. 
This latter condition is essential, so it is fairly easy to break Hullot's lifting lemma by dropping it\longtxt{: e.g.\ under the term rewriting system (\trs) $\{f(0,1) \tor 2, \coin \tor 0, \coin \tor 1\}$ the term rewriting derivation  $f(X,X)[X/\coin] \rw^* 2$ cannot be lifted by any narrowing derivation}. 
%
Several variants and extensions of narrowing have been developed in order to improve that result under certain assumptions or for particular classes of term rewriting systems~\cite{DBLP:journals/aaecc/MiddeldorpH94,MeseguerT07Reach,DBLP:conf/rta/DuranEEMT11}.

\medskip

In this paper we show how to adapt the generator-based extra variable elimination transformation used 
in functional-logic programming (FLP) to drop the normalization condition required by Hullot's lifting
lemma. The proposed technique is devised for left-linear constructor systems (\ctrss)
with extra variables, and it is limited to derivations reaching a ground expression. 
%
%
To test the feasibility of this approach, we have also developed a prototype in Maude~\cite{maude-book}, 
relying on the natural rewriting on-demand strategy~\cite{escobar04} to obtain an effective operational procedure.

The rest of the paper is organized as follows. In Section \ref{sect:setting} we introduce the semantics
for \ctrss\ that we have used to formally prove the results, and that first suggested us the feasibility
of the approach. In Section \ref{sect:generators} we show our adaptation of the generators technique from
FLP, and use the semantics for proving the adequacy of the technique for lifting term rewriting derivations
reaching ground c-terms. In Section \ref{maude} we outline the implementation and commands of our prototype.
Finally Section \ref{sect:conclusions} concludes and outlines some lines of future work.

\section{Prelimininaries and formal setting}\label{sect:setting}

We mostly use the notation from~\cite{baader-nipkow}, with some additions from~\cite{Lopez-FraguasRS09-RTA09}. 
We consider a first order signature $\Sigma = \CS \uplus \FS$, where $\CS$ and $\FS$ are two disjoint
sets of \emph{constructor} and defined \emph{function} symbols respectively, all of them with associated
arity. 
We use $c,d,\ldots$ for constructors, $f,g,\ldots$ for functions and $X,Y,\ldots$ for variables of a
numerable set  $\var$. The notation $\overline{o}$ stands for tuples of any kind of syntactic objects. 
The set $\Exp$ of {\it total expressions} is defined as $\Exp \ni e::= X \mid h(e_1,\ldots,e_n)$, where 
$X\in\var$, $h\in \CS^n\cup \FS^n$ and $e_1,\ldots,e_n\in \Exp$. The set $\CTerm$ of
{\it total constructed terms}
(or {\it  c-terms}) is defined like $\Exp$, but with $h$ restricted to $\CS^n$ (so $\CTerm\subseteq \Exp$).
The intended meaning is that $\Exp$ stands for evaluable expressions, i.e., expressions that can contain 
function symbols, while $\CTerm$ stands for data terms representing {values}. 
We will write $e,e',\ldots$ for expressions and $t,s,\ldots$ for c-terms. 
We say that an expression $e$ is \emph{ground} iff no variable appears in $e$. 
We will frequently use \emph{one-hole contexts}, defined as $\Cntxt\ni {\cal C} ::= [\ ] \mid h(e_1,\ldots,{\cal C},\ldots,e_n)$. 

\begin{example}\label{ex:clerks}
We will use a simple example throughout this section to illustrate these definitions. Assume we want to represent
the staff of a shop, so we have $\CS = \{\mi{madrid}^0, \mi{vigo}^0,$
$\mi{man}^0, \mi{woman}^0, \mi{pepe}^0, \mi{luis}^0,$ $\mi{pilar}^0,$ $\mi{maria}^0, e^2, p^2\}$,
where $e$ will be the constructor for employees and $p$ the constructor for pairs, and
$\FS = \{\mi{branches}^0, \mi{search}^1, \mi{employees}^1\}$.
Using this signature, we can build the set $\Exp = \{\mi{madrid}, \mi{vigo},$ $\mi{employees}$ $(\mi{madrid}),
p(\mi{pilar}, X), \ldots\}$. From this set, we have $\CTerm = \{\mi{madrid}, \mi{vigo},
p(\mi{pilar}, X), \ldots\}$,
while the ground terms are $\{\mi{employees}(\mi{madrid}),$ $\mi{madrid}, \mi{vigo}, \ldots\}$.
Finally, a possible one-hole context is $p([\ ], X)$.
\end{example}

We also consider the extended signature $\Sigma_\bot
=\Sigma\cup \{\bot\}$, where $\bot$ is a new $0$-arity constructor symbol that 
does not appear in programs and which stands for the undefined value. Over this signature 
we define the sets $\Exp_\bot$ and $\CTerm_\bot$ of {\it partial} expressions and c-terms, respectively.
The intended meaning is that $\Exp_\bot$ and $\CTerm_\bot$ stand for 
partial expressions and values, respectively.  
Partial expressions are ordered by the {\em approximation} ordering $\sqsubseteq$ defined as the least
partial ordering satisfying $\perp \sqsubseteq e$ and $e \sqsubseteq e'
\Rightarrow {\cal C}[e] \sqsubseteq {\cal C}[e']$
for all $e,e' \in \Exp_\perp, {\cal C} \in {\it \Cntxt}$.
The {\it shell} $|e|$ of an expression $e$ represents the outer constructed part
of $e$ and is defined as: $|X|=X$; $|c(e_1,\ldots,e_n)|=c(|e_1|,\ldots,|e_n|)$; $|f(e_1,\ldots,e_n)|=\bot$. \longtxt{It is trivial to check that for any expression $e$ we have $|e| \in \CTerm_\perp$, that any total expression is maximal w.r.t.\ $\ordap$, and that as consequence if $t$ is total then $t \ordap |e|$ implies $t = e$.} 

\begin{example}
Using the signature from Example~\ref{ex:clerks}, we have
$\mi{employees(\bot)} \in \Exp_\bot$,
$p(\bot, X) \in \CTerm_\bot$, and
$|p(\mi{search}(\mi{branches}), X)| = p(\bot, X)$.
\end{example}

{\em Substitutions} $\theta \in \Subst$ are finite mappings $\theta:\var
\longrightarrow \Exp$, extending naturally to $\theta:\Exp \longrightarrow
\Exp$.  \longtxt{We write $\epsilon$ for the identity (or empty) substitution.} 
We write $e\theta$ to apply of $\theta$ to $e$, and $\theta\theta'$
for the composition, defined by $X(\theta\theta') = (X\theta)\theta'$. 
\longtxt{The domain and  variable range of $\theta$ are defined as $\dom(\theta) = \{X\in \var \mid X\theta \neq X\}$ and $\vran(\theta) = \bigcup_{X\in \mi{dom}(\theta)}\mi{var}(X\theta)$.}
\shorttxt{The domain of $\theta$ is defined as $\dom(\theta) = \{X\in \var \mid X\theta \neq X\}$.}
By $[X_1/e_1, \ldots, X_n/e_n]$ we denote a substitution $\sigma$ such that $\dom(\sigma) = \{X_1, \ldots, X_n\}$ and $\forall i. \sigma(X_i) = e_i$.  
\longtxt{If $\dom(\theta_0) \cap \dom(\theta_1) = \emptyset$, their disjoint union $\theta_0 \uplus \theta_1$ is defined by $(\theta_0 \uplus \theta_1)(X) = \theta_i(X)$, if $X \in \dom(\theta_i)$ for some $\theta_i$; $(\theta_0 \uplus \theta_1)(X) = X$ otherwise.}
\longtxt{Given $W \subseteq {\cal V}$
we write 
$\theta|_{W}$ for the restriction of $\theta$ to $W$, i.e.\ $(\theta|_{W})(X) = \theta(X)$ if $X \in W$, and $(\theta|_{W})(X) = X$ otherwise; we use $\theta|_{\backslash D}$ as a shortcut for $\theta|_{({\cal V} \backslash D)}$.} 
\emph{C-substitutions}
$\theta \in \emph{CSubst}$ verify that $X\theta \in \CTerm$ for all $X\in \dom(\theta)$. \longtxt{We say a substitution $\sigma$ is ground  iff $\vran(\sigma) = \emptyset$, i.e.\ $\forall X \in \dom(\sigma)$ we have that $\sigma(X)$ is ground.}\shorttxt{We say that a substitution is ground when no variable appears in its range.} The sets $\Subst_\perp$ and $CSubst_\perp$ of partial substitutions and partial c-substitutions are the sets of finite mappings from variables to partial expressions and partial c-terms, respectively. 

\begin{example}
Using the signature from Example~\ref{ex:clerks}, we can define the C-substitutions
$\theta_1 \equiv X / \mi{woman}$, $\theta_2 \equiv X / \mi{man}$,
and $\theta_3 \equiv Y / \mi{pilar}$. We can define the restrictions 
$\theta_1|_{\{X\}} = \theta_1$ and
$\theta_1|_{\backslash \{X\}} = \epsilon$.
Finally, given the expression $p(X, Y)$ we have
$p(X, Y)\theta_1\theta_2 = p(\mi{woman}, Y)$
and
$p(X, Y)\theta_1\theta_3 = p(X, Y)\theta_3\theta_1 = p(\mi{woman}, \mi{pilar})$.
\end{example}

A {left-linear constructor-based term rewriting system} or just \emph{constructor system} (\emph{\ctrs}) or \emph{program} ${\cal P}$ is a set of c-rewrite rules of the form $f(\overline{t})\to r$ where $f\in \FS^n$, $r\in \Exp$ and $\overline{t}$ is a linear $n$-tuple of c-terms, where linearity means that variables occur only once in $\overline{t}$. Notice that we allow $r$ to contain so called \emph{extra variables}, i.e., variables not occurring in $f(\overline{t})$. 
\longtxt{To be precise, we say that $X \in \var$ is an extra variable in the rule $l \tor r$
iff $X \in \mi{var}(r) \setminus \mi{var}(l)$, and by} \shorttxt{By} $\vextra{R}$ we denote
the set of extra variables in a program rule $R$. 
 We assume that every \ctrs\ 
contains the rules $\progq = \{X~?~Y \tor X, X~?~Y \tor Y\}$, defining the behavior of $? \in \FS^2$, used in infix mode\shorttxt{ and right-associative}, and that those are the only rules for $?$. \longtxt{Besides, $?$ is right-associative so $e_1~?~e_2~?~e_3$ is equivalent to $e_1~?~(e_2~?~e_3)$.} For the sake of conciseness we will often omit\longtxt{ these rules when presenting a \ctrs}.  A consequence of this is that we only consider non-confluent programs. 
Given a \trs\ ${\cal P}$, its associated \emph{term rewriting relation} $\rw_{\cal P}$ is defined as:
${\cal C}[l\sigma] \rw_{\cal P} {\cal C}[r\sigma]$
for any context ${\cal C}$, rule $l \tor r \in {\cal P}$ and $\sigma \in
\emph{Subst}$. 
We write $\stackrel{*}{\rw_{\cal P}}$ for the reflexive and
transitive closure of the relation $\rw_{\cal P}$. 
We will usually omit the reference to ${\cal  P}$ or denote it by $\prog \vdash e \rw e'$ and $\prog \vdash e \rw^* e'$.  

\begin{example}\label{ex:rules}
Using the signature from Example~\ref{ex:clerks}, we can describe the following
program:
$$
\begin{array}{lll}
  \mi{branches} & \to & \mi{madrid}\ ?\ \mi{vigo}\\
  \mi{employees}(\mi{madrid}) & \to & e(\mi{pepe}, \mi{men})\\
  \mi{employees}(\mi{madrid}) & \to & e(\mi{maria}, \mi{men})\\
  \mi{employees}(\mi{vigo}) & \to & e(\mi{pilar}, \mi{women})\ ?\ e(\mi{luis}, \mi{men})\\
  \mi{search}(e(N,S)) & \to & p(N,N)
\end{array}
$$
In this example, the function symbol $\mi{branches}$ defines the different branches of the
company, $\mi{employees}$ defines the employees in each branch (built with the constructor
symbol $e$), and $\mi{search}$ returns a pair of names, built with the constructor symbol $p$.
Note that several different notations are possible; for example, it is possible to define
the employees of one branch by using just one rule and the $?$ operator or just several
different rules with the same lefthand side. 
\end{example}

%

\subsection{A proof calculus for constructor systems with extra variables}
In~\cite{Lopez-FraguasRS09-RTA09} an adequate semantics for reachability of c-terms by term rewriting in
\ctrss\ was presented. The key idea there was using a suitable notion of value, in this case
the notion of s-cterm.
$\scterm$ is the set of s-cterms, which are 
\emph{finite} sets of elemental s-cterms, 
while the set $\escterm$ of elemental s-cterms is defined as 
$\escterm \ni e\uscterm{} ::= X~|~c(\uscterm{1}, \ldots, \uscterm{n})$ for $X \in \var$, $c \in\cs^n$, 
$\uscterm{1}, \ldots, \uscterm{n} \in \scterm$. We extend this \longtxt{idea} to expressions obtaining 
the sets $\sexp$ of s-expressions or just s-exp, and $E\sexp$ of elemental s-expressions, which are defined 
the same but now using any symbol in $\Sigma$ in applications instead of just constructor symbols. Note 
that the s-expression $\emptyset$ corresponds to $\perp$, so s-exps are partial by default. The approximation
preorder $\sqsubseteq$ is defined for s-exps as the least preorder such that $\mi{se}\sqsubseteq \mi{se}'$
iff $\forall \ese\in \mi{se}.\exists \ese'\in \mi{se}'$ such that $\ese\sqsubseteq \ese'$, $X\sqsubseteq X$
for any $X\in\var$, and $h(\mi{se}_1, \ldots, \mi{se}_n)\sqsubseteq h(\mi{se}'_1, \ldots, \mi{se}'_n)$ 
iff $\forall i. \mi{se}_i\sqsubseteq \mi{se}'_i$.

\begin{example}
Using the signature from Example~\ref{ex:clerks}, and given the s-cterm
$\mi{sct} \equiv e(\{\mi{pepe}, \mi{pilar}\},$ $\{\mi{men},$ $\mi{women}\})$,
we have $\mi{sct} \in \escterm$, while $\{\mi{sct}\} \in \scterm$.
Similarly, given the es-exp
$\mi{esex} \equiv \mi{employees}$ $(\{\mi{madrid}, \mi{vigo}\})$ we have $\mi{esex} \in \mi{E\sexp}$
and $\mi{esex} \not\in \escterm$. Finally, we have that $\{\mi{esex}\} \in \sexp$.
\end{example}

The sets $\ssubst$ and $\scsubst$ of s-substitutions and s-csubstitutions (or just s-csubst) \longtxt{consist of finite mappings from variables to s-exps or s-cterms, respectively}\shorttxt{are substituions with s-exps or s-cterms in their range, respectively}.  
\longtxt{We extend  s-substs to be applied to $\mi{ESExp}$ and $\mi{SExp}$ as 
$\sigma : \mi{E\sexp} \rightarrow \sexp$ defined by }\shorttxt{To apply  s-substs to s-exp we use }
$X\sigma = \sigma(X)$, $h(\overline{se})\sigma  =  \{h(\overline{se\sigma})\}$\longtxt{; and $\sigma : \sexp \rightarrow \sexp$ defined by }\shorttxt{, }$\mi{se}\sigma =  \bigcup_{\ese \in \mi{se}} \ese\sigma$. \longtxt{The approximation preorder $\ordap$ is defined for s-substs as $\sigma \sqsubseteq \theta$ iff $\forall X \in \var. \sigma(X) \sqsubseteq \theta(X)$.
 For any nonempty and finite set $\{\theta_1, \ldots, \theta_n\} \subseteq \scsubst\sbt$ we define $\bigcup \{\theta_1, \ldots, \theta_n\} \in \scsubst\sbt$ as $\bigcup \{\theta_1, \ldots, \theta_n\}(X) = \theta_1(X) \cup \ldots \cup \theta_n(X)$.}
 
\begin{example}
Using the signature from Example~\ref{ex:clerks}, we can define the s-csubstitution
$\sigma \equiv \{ X / \{\mi{pepe}, \mi{pilar}\},$
$Y / \{\mi{men}, \mi{women}\}\} \in \scsubst$. Hence, given
$\mi{esex} \equiv e(\{X\}, \{Y\}) \in \mi{E\sexp}$ we have that
$\mi{esex}\sigma \equiv e(\{\mi{pepe},$ $\mi{pilar}\},$ $\{\mi{men}, \mi{women}\})$.
\end{example}

\longtxt{We obtain the denotation of an expression  as }\shorttxt{The denotation of an expression is }the denotation of its associated s-expression, assigned by the operator $\etose{\_} : \mi{Exp}_\perp \tor \sexp$, 
 defined as $\etose{\bot}=\emptyset$; $\etose{X} = \{X\}$ for any $X\in\var$; $\wetose{h(e_1, \ldots, e_n)} = \{h(\etose{e_1}, \ldots, \etose{e_n})\}$ for any $h\in\Sigma^n$\longtxt{. The operator $\etose{\_}$ is}\shorttxt{, } extended to s-substitutions as $\etose{\sigma}(X)=\wetose{\sigma(X)}$, for $\sigma \in Subst_\perp$. \longtxt{It is easy to check that $\wetose{e\sigma} = \etose{e}\etose{\sigma}$ (see~\cite{Lopez-FraguasRS09-RTA09}).} Conversely, we can flatten \longtxt{an s-expression $se$} \shorttxt{s-exps }to obtain the set $\miflat(se)$ of expressions ``contained'' in it, so $\miflat(\emptyset)=\{\bot\}$\longtxt{ and }\shorttxt{, }$\miflat(se)=\bigcup_{\ese\in se} \miflat(\ese)$ if $se\not=\emptyset$, \longtxt{ where the flattening of elemental s-exps is defined as} $\miflat(X)=\{X\}$\longtxt{ ; }\shorttxt{, }$\miflat(h(se_1,\ldots,se_n))=\{h(e_1,\ldots,e_n)~|~e_i\in \miflat(se_i)\mbox{ for } i=1..n\}$. 
%
%
\longtxt{\begin{figure}[tb]
\begin{center}
\framebox{
\begin{minipage}{.9\textwidth}
\begin{center}
\begin{tabular}{l@{~~}cl}
 \textbf{\crule{E}} & $\usexp{} \cltor \emptyset$ \\[.15cm]
 \textbf{\crule{RR}} & $\{X\} \cltor \{X\}$ & if $X \in \var$ \\[.15cm]
\textbf{\crule{DC}} & $\infer[]{\{c(\usexp{1}, \ldots, \usexp{n})\} \cltor \{c(\uscterm{1}, \ldots, \uscterm{n})\}}{\usexp{1} \cltor \uscterm{1} \ \ldots \ \usexp{n} \cltor \uscterm{n}}$ & if $c \in \CS$ \\[.15cm]
\textbf{\crule{More}} & $\infer[]{\usexp{} \cltor \uscterm{1} \cup \ldots \cup \uscterm{n}}{\usexp{} \cltor \uscterm{1} \ \ldots \usexp{} \cltor \uscterm{n}}$\\[.15cm]
\textbf{\crule{Less}} & $\infer[]{\{e\usexp{1}, \ldots, e\usexp{n}\} \cltor \uscterm{1} \cup \ldots \cup \uscterm{m}}{\{\mi{esa}_1\} \cltor \uscterm{1} \ \ldots \ \{\mi{esa}_m\} \cltor \uscterm{m}}$ & $\begin{array}{l}
\mbox{if } n \geq 2, m>0 \mbox{, for any }\\
\{\mi{esa}_1, \ldots, \mi{esa}_m\}  \\
\subseteq \{e\usexp{1}, \ldots, e\usexp{n}\}
\end{array}
$\\[.15cm]
 \textbf{\crule{ROR}} & $\infer[]{\{f(se_1,\ldots,se_n)\} \cltor st}{se_1 \cltor \etose{p_1}\theta \ \ldots \ se_n \cltor \etose{p_n}\theta \ ~\etose{r}\theta \cltor st}$ & if $\begin{array}{l}
(f(p_1, \ldots, p_n) \tor r)\in \prog \\
\theta \in \scsubst\sbt
\end{array}$
\end{tabular}
\end{center}
\end{minipage}
}
\end{center}
\vspace{-.5cm}
    \caption{A proof calculus for constructor systems}
    \label{fig:semCRW}
\end{figure}}
\shorttxt{The semantics of s-expressions is defined by a sequent calculus that proves reduction statements of the form $\prog \vdash se \clto st$ with $se \in \sexpb$ and $st \in \sctermb$, expressing that $st$ represents an approximation to one of the possible structured sets of values for $se$ under the $\ctrs$ $\prog$, with the rules \textbf{\crule{(E)}}  $\usexp{} \cltor \emptyset$; 
\textbf{\crule{(RR)}} $\{X\} \cltor \{X\}$ if $X \in \var$;
\textbf{\crule{(DC)}} $\{c(\usexp{1}, \ldots, \usexp{n})\} \cltor \{c(\uscterm{1}, \ldots, \uscterm{n})\}$ if $\forall i \in \{1, \ldots, n\}. \usexp{i} \cltor \uscterm{i}$ for $c \in \CS$; 
\textbf{\crule{(More)}}  ${\usexp{} \cltor \uscterm{1} \cup \ldots \cup \uscterm{n}} $ if $\forall i \in \{1, \ldots, n\}. \usexp{} \cltor \uscterm{i}$;
\textbf{\crule{(Less)}} $\{e\usexp{1}, \ldots, e\usexp{n}\} \cltor \uscterm{1}$ $\cup$ $\ldots \cup \uscterm{m}$ if $\forall i \in \{1, \ldots, m\}. \{\mi{esa}_i\} \cltor \uscterm{i}$ with $n \geq 2, m>0$ for any $\{\mi{esa}_1, \ldots, \mi{esa}_m\} \subseteq \{e\usexp{1}, \ldots, e\usexp{n}\}$; 
\textbf{\crule{(ROR)}} ${\{f(se_1,\ldots,se_n)\} \cltor st}$ if $\forall i \in \{1, \ldots, n\}. se_i \cltor \etose{p_i}\theta \wedge \etose{r}\theta \cltor st$ for $(f(p_1, \ldots, p_n) \tor r)\in \prog$ and  $\theta \in \scsubst\sbt$.}  

\begin{example}
Using the signature from Example~\ref{ex:clerks}, we have that $\etose{p(X, Y)} = \{p(\{X\}, \{Y\})\}$ and
$\mi{flat}(\{p(\{X\},$ $\{Y, Z\})\}) = \{p(X, Y), p(X, Z)\}$
\end{example}

\longtxt{In Figure \ref{fig:semCRW} we can find the proof calculus that defines the semantics of s-expressions.  Our proof calculus proves reduction statements of the form $se \clto st$ with $se \in \sexpb$ and $st \in \sctermb$, expressing that $st$ represents an approximation to one of the possible structured sets of values for $se$.} 
\longtxt{We refer the interested reader to~\cite{Lopez-FraguasRS09-RTA09} for detailed explanations about the calculus.}\longtxt{We write $\prog \vdash se \clto st$ to express that $se \clto st$ is derivable in our calculus under the \ctrs\ $\prog$.} \longtxt{We say that a proof for a statement $\prog \vdash se \clto st$ is ground iff $se$, $st$ and all the s-exp in the premises are ground.} The \emph{denotation} of an s-expression $se$ under a \ctrs\ $\prog$ is defined as $\den{se}^{\prog} = \{st \in \sctermb~|~\prog \vdash se \clto st\}$, so ${\den{e}}^{\prog} = {\den{\etose{e}}}^{\prog}$. \longtxt{In the following w}\shorttxt{W}e will usually omit the reference to $\prog$. The denotation of $\sigma \in \ssubst$ is defined as $\den{\sigma} = \{\theta \in \scsubst\sbt~|~ \forall X \in \var, \sigma(X) \clto \theta(X)\}$, so for $\theta \in Subst_\perp$ we define $\den{\theta} = \den{\etose{\theta}}$.

\begin{example}
Using the signature from Example~\ref{ex:clerks} and the rules from Example~\ref{ex:rules}, we have 
$\mi{employees}(\{\mi{X}\})$ $\clto \{e(\mi{pepe}, \mi{men})\}$, given the substitution
$X / \{\mi{madrid}\}$. Moreover, we can use this same substitution to reach
$\{e(\mi{maria}, \mi{men})\}$ by using a different program rule.
\end{example}
  
\shorttxt{In~\cite{Lopez-FraguasRS09-RTA09} we proved the adequacy of the calculus w.r.t.\ term rewriting for programs without extra variables. As extra variables are very common when using narrowing, we have extended the proof to deal with them. A remarkable point is that, as a consequence of the freely instantiation of extra variables performed in \textbf{\crule{ROR}}, then every program with extra variables turns into non-deterministic. For example under a program $\{f \tor (X, X)\}$ and using $0, 1 \in \CS^0$ we can prove $\etose{f} = \{f\} \clto \{(\{0\}, \{1\})\} = \etose{(0,1)}$.} 
\longtxt{The setting presented in~\cite{Lopez-FraguasRS09-RTA09} was not able to deal with extra variables. As programs with extra variables are very common when using narrowing, 
for this work we decided to extend 
the setting to deal with them. But then we realized that the semantics had the foundations
to deal with extra variables, 
as the rule \textbf{\crule{ROR}} from Figure \ref{fig:semCRW} 
allows to instantiate extra variables freely with s-cterms: therefore all that was left was proving the adecuacy of the semantics in this extended scenario.
Nevertheless, as a consequence of 
the freely instantiation of extra variables in \textbf{\crule{ROR}}, then
every program with extra variables turns into non-deterministic. For example consider a program $\{f \tor (X, X)\}$ for which the constructors $0, 1 \in \CS^0$ are available, then we can do:
$$
\begin{scriptsize}
\infer[\textbf{\crule{ROR}}]{\etose{f} = \{f\} \clto \{(\{0\}, \{1\})\} = \etose{(0,1)}}
{
\infer[\textbf{\crule{DC}}]{\etose{(X, X)}[X/\{0,1\}] = \{(\{0,1\}, \{0,1\})\} \clto \{(\{0\}, \{1\})\}}
{
\infer[\textbf{\crule{Less}}]{\{0,1\}  \clto \{0\}}
		{\infer[\textbf{\crule{DC}}]{\{0\} \clto \{0\}}{}} &
\infer[]{~\{0,1\}  \clto \{1\}}{\ldots}
 }
}
\end{scriptsize}
$$}

But in fact this is not very surprising, and it has to do with the relation between non-determinism and
extra variables~\cite{AntoyH06Extra}, but adapted to the run-time choice semantics~\cite{hussmann93,rodH08}
induced by term rewriting.  
As a consequence of this we assume that all the programs contain the function $?$, so we only consider 
non-confluent \trss. We admit that this is a limitation of our setting, but we also conjecture that 
for confluent \trss\ a simpler semantics could be used, for which the packing of alternatives of c-terms
would not be needed. However, the important point to bear in mind is that
having $?$ at one's disposal is enough to express the non-determinism of any program~\cite{Han05TR}, so
we can use it to define the transformation $\setoe{\_}$ from s-exp and elemental s-exp to partial 
expressions that, contrary to $\miflat$, now takes care of the keeping the nested set structure by means 
of uses of the $?$ function. Then $\setoe{\_} : \esexp \rightarrow \Exp_\perp$ is defined by 
$\setoe{X} = X$, $\setoe{h(se_1, \ldots, se_n)} = h(\setoe{se_1}, \ldots, \setoe{se_n})$; and 
$\setoe{\_} : \sexp \rightarrow \Exp_\perp$ is defined by $\setoe{\emptyset} = \perp$, 
$\setoe{\{ese_1, \ldots, ese_n\}} = \setoe{ese_1}~?~\ldots~?~\setoe{ese_n}$ for $n > 0$, where in the 
case for $\setoe{\{ese_1, \ldots, ese_n\}}$ we use some fixed arbitrary order on terms in the line of 
Prolog~\cite{SterlingShapiro86} for arranging the arguments of $?$. This operator is also overloaded 
for substitutions as $\setoe{\_} : \ssubst \rightarrow \Subst_\perp$ as 
$(\setoe{\sigma})(X) = \setoe{\sigma(X)}$.
Thanks to the power of $?$ to express non-determinism, that transformation preserves the semantics from
Figure~\ref{fig:semCRW}, and we can use it to prove the following new result about the adequacy of the
semantics for programs with extra variables---see~\cite{generatorsProofs} for a detailed proof.

\begin{theorem}[Adequacy of $\den{\_}$] 
For all $e, e' \in \Exp, t \in \CTerm_\perp, st \in \sctermb$: \\
\textbf{Soundness }$st \in \den{\etose{e}}$ and $t \in \miflat(st)$ implies $e \rw^* e'$ for some $e' \in \Exp$ such that $t \ordap |e'|$. Therefore, $\etose{t} \in \den{\etose{e}}$ implies 
$e \rw^* e'$ for some $e' \in \Exp$ such that $t \ordap |e'|$. Besides, in any of the previous cases, if $t$ is total then $e \rw^* t$.\\
\textbf{Completeness} $e \rw^* e'$ implies $\wetose{|e'|} \in \den{\etose{e}}$. Hence, if $t$ is total then $e \rw^* t$ implies $\wetose{t} \in \den{\etose{e}}$. 
\end{theorem}
\longtxt{
\begin{figure}[bt]
\begin{center}
\framebox[\textwidth]{
\begin{tabular}[t]{c|c}
$
\begin{array}{l}
\_  \vdash \mrel{\_}{\_} \subseteq \ctrs \times \sctermb \times \Exp \\
\begin{array}{ll}
\prog \vdash \mrel{st}{e} & \mbox{ if } \forall est \in st, \prog \vdash \mrel{est}{e} \\
\end{array}
\; \\ \; \\ \; \\
\end{array}
$
 &
$
\begin{array}{l}
\_ \vdash \mrel{\_}{\_} \subseteq \ctrs \times \esctermb \times \Exp \\
\begin{array}{ll}
\prog \vdash \mrel{X}{e} & \mbox{ if } \prog \vdash e \rw^* X \\
\prog \vdash \mrel{c(\overline{st})}{e} & \mbox{ if } \prog \vdash e \rw^* c(\overline{e})  \mbox{ for some } \overline{e}
\end{array}\\
\mbox{ ~~~~~~~~~~~~~~~such that } \forall e_i \in \overline{e}, \prog \vdash \mrel{st_i}{e_i}\\
\end{array}
 $
\end{tabular}
}
 \end{center}
\vspace{-.5cm}
    \caption{Domination relation}
    \label{fig:DomRel}
\end{figure}
}
We refer the interested reader to~\cite{Lopez-FraguasRS09-RTA09} and~\cite{LRS09ReportFully}
(Theorems 2 and 3) for more properties of $\den{\_}$ like compositionality or monotonicity, some of which are used in the
proofs for the results in the present paper.
\longtxt{There is another characterization of $\den{\_}$ closer to term rewriting which is based of the  \emph{domination relation $\mrel{\_}{\_}$} presented in Figure~\ref{fig:DomRel} (we will omit the prefix ``$\prog \vdash$'' when it is implied by the context).}
\longtxt{With this relation we try to transfer to the rewriting world the finer distinction between sets of values that the structured representation of $\sctermb$ allows us to perform.  We extend the relation $\mrel{\_}{\_}$ to $\_\vdash \mrel{\_}{\_} \subseteq \ctrs \times \scsubst\sbt \times Subst$ by $\mrel{\theta}{\sigma}$ iff $\forall X \in \var, \mrel{\theta(X)}{\sigma(X)}$. As can be seen in~\cite{LRS09ReportFully}, this relation is a key ingredient to prove the soundness of $\den{\_}$, and its equivalence to $\den{\_}$ is stated in the following result. 
\begin{lemma}[Domination]\label{TMrelvsSem}\label{lemDenSubstMrel1}
For all $ e \in \Exp, \mi{st} \in \sctermb$, $st \in \den{\etose{e}}$ iff $\mrel{st}{e}$. Besides, regarding substitutions, for all $\sigma \in \Subst$, $\theta \in \scsubst$ we have that $\theta \in \den{\etose{\sigma}}$ iff $\mrel{\theta}{\sigma}$.
\end{lemma}}

\section{The generators approach}\label{sect:generators}

\longtxt{In this section we will show a proposal for adapting the generators technique from the field of functional-logic programming \cite{DiosLopez07,AntoyH06Extra} to the lifting of term rewriting derivations from arbitrary instances of expressions. This technique consists in replacing free and extra variables by a call to a \emph{generator function} that can be reduced to any ground c-term.}%
\shorttxt{The generators technique from the field of FLP \cite{DiosLopez07,AntoyH06Extra} consists in replacing free and extra variables by a call to a \emph{generator function} that can be reduced to any ground c-term.} The generator function $\gen$ is defined as follows:
\begin{definition}[Generator function]\label{def:generator}
For any program $\prog$ we can define a fresh function $\gen$ as follows: for each $c \in \CS^n$ we add a new rule $\gen \tor c(\gen, \ldots, \gen)$ to the program. By $\progg$ we denote the program that consists of the set of rules for $\gen$.
\end{definition}

\begin{example}
Given the system in Example~\ref{ex:rules}, the rules for $\gen$ are $\progg \equiv \{\gen \to \mi{madrid},
\gen \to \mi{vigo}, \gen \to \mi{pepe}, \gen \to \mi{luis}, \gen \to \mi{maria}, \gen \to \mi{pilar},
\gen \to \mi{men}, \gen \to \mi{women},$ $\gen \to e(\gen,$ $\gen),$ $\gen \to p(\gen, \gen)\}$.
\end{example}

The point with $\gen$ is that 
we can use it to compute any \emph{ground value}: 
\begin{proposition}\label{thGenCompAux1}
For all $t \in \CTerm$, $st \in \scterm$ and $\theta \in \scsubst$ such that those are ground we have $\gen \rw^* t$, $st \in \den{\gen}$ and $\theta \in \den{[\overline{X/gen}]}$ for $\overline{X} = \dom(\theta)$.
\end{proposition}

Then the main idea with generators is that given some $e \in \Exp$ with $\mi{var}(e) = \overline{X}$, we can simulate narrowing with $e$ by performing term rewriting with $e[\overline{X/gen}]$.
As $gen$ can be reduced to any ground s-cterm, then 
Lemma 1 from \cite{Lopez-FraguasRS09-RTA09} 
suggests that
this procedure will be able to lift derivations $e\sigma \rw^* t$ with an arbitrary $\sigma \in Subst$\longtxt{, even those which are not normalized: e.g. we can easily apply this technique to 
the example in Section~\ref{sect:intro}, getting $f(X,X)[X/gen] \rw^* f(0,1) \rw 2$}. 
Sadly, on the other hand, only derivations reaching a ground c-term will be lifted, and the reason for that is that $gen$ can be reduced to an arbitrary ground c-term, but it cannot be reduced to any c-term with variables. Thus, under the program $\{g(c(X)) \tor X\}$ the term rewriting derivation $g(Y)[Y/c(X)] \rw X$ cannot be lifted by using generators, as $g(Y)[Y/gen] \rw g(c(gen)) \rw gen \not\rw^* X$, even though $[Y/c(X)]$ is a normalized substitution.

In order to prove the completeness of the generators technique for the reachability of ground c-terms, we rely on the following modification of Lemma 1 from \cite{Lopez-FraguasRS09-RTA09}. 

\begin{lemma}\label{thGenCompAux2}
For all $\sigma \in \ssubst$, $\mi{se} \in \sexp$, $\mi{st} \in \scterm$, if $\mi{st}$ is ground then $\mi{se}\sigma \clto \mi{st}$ implies $\exists \theta \in \den{\sigma}$ such that $\mi{se}\theta \clto \mi{st}$, $\theta$ is ground and $\dom(\theta) = \dom(\sigma)$.
\end{lemma}

Note the restriction to ground s-cterms in Lemma \ref{thGenCompAux2} is crucial, and that it reflects the 
lack of completeness for reaching non-ground c-terms of the generators technique: e.g. under the program $\{f \tor c(X)\}$ using $se = \{Y\}$, $\sigma = [Y/\{f\}]$ and $st = \{c(\{X\})\}$ the only $\theta \in \den{[Y/\{f\}]}$ fulfilling the first condition is $\theta = [Y/\{c(\{X\})\}]$, which is not ground. On the other hand those s-csubst obtained by Lemma \ref{thGenCompAux2}  are ground, and so they are in the denotation of an appropriate substitution with only generators in its range.

\bigskip
Generators can be introduced in programs systematically in order to eliminate extra variables from program
rules using a program transformation in the line of \longtxt{those from} \cite{DiosLopez07,AntoyH06Extra}.
In those works the usual call-time choice semantics for functional-logic programming~\cite{GHLR99} was
adopted, therefore we use a different transformation that is adapted to the use of term rewriting, 
which leads to a different 
set of reachable c-terms \longtxt{than that obtained with call-time choice}~\cite{rodH08}. The point in eliminating extra variables is that in this way we eliminate the ``oracular guessing'' that is performing in a term rewriting step using extra variables: \longtxt{by this guessing we refer for example to}\shorttxt{for example} the instantiation performed under the program $\{f \tor g(X), g(0) \tor 1\}$ in the first step of the derivation $f \rw g(0) \rw 1$ for the extra variable $X$\longtxt{, that has to be instantiated with $0$ in order for the derivation to continue}.
That, combined with a suitable on-demand evaluation strategy like natural rewriting \cite{escobar04}, turns term rewriting with generators into an effective 
mechanism for lifting term rewriting derivations. \longtxt{We formalize our extra variable elimination transformation through the following definition.}

\begin{definition}[Generators program transformation]~\\
Given a program $\prog$ its transformation $\gt{\prog}$ consists of the rules $\progg$ for $\mi{gen}$ together with the transformation of each rule in $\prog$, defined as $\wgt{f(p_1, \ldots, p_n) \tor r} = f(p_1, \ldots, p_n) \tor r[\overline{X/gen}]$, where $\overline{X} = \vextra{f(p_1, \ldots, p_n) \tor r}$.
\end{definition}
Then it is clear that for any program $\prog$ its transformation $\gt{\prog}$ does not have any extra variable in its rules. \longtxt{Note that, c}\shorttxt{C}ontrary to the proposals
from~\cite{DiosLopez07,AntoyH06Extra},  \longtxt{this transformation destroys}\shorttxt{we destroy} the sharing that normally appears when there are several occurrences of the same variable, in procedures that instantiate variables like narrowing or SLD resolution. In our transformation, however, once instantiated with $gen$ every occurrence of the same variable evolves independently. This is needed to ensure completeness under the transformed program, which can be seen considering the program $\prog = \{f \tor (g(X), h(X), g(0) \tor 1, h(1) \tor 2\}$ and the derivation $\prog \vdash f \rw (g(0~?~1), h(0~?~1)) \rw^* (g(0), h(1)) \rw^* (1,2)$: as extra variables can be instantiated with arbitrary expressions that implies that in particular those can be instantiated with ``alternatives'' of expressions built using the $?$ function, which can evolve independently after the alternative between them is resolved. We can lift that derivation with our transformation as $\gt{\prog} \vdash f \rw (g(gen), h(gen)) \rw^* (g(0), h(1)) \rw^* (1,2)$.
%
%
The adequacy of the transformation is formulated in the following result, in the same terms as the variable elimination result from~\cite{DiosLopez07}.
\begin{theorem}\label{th:adeqGenTr}
For any program $\prog$, $se \in \mi{SExp}$, $st \in \mi{SCTerm}$ if $st$ is ground then $\progg \uplus \prog \vdash se \clto st$ iff $\gt{\prog} \vdash se \clto st$. 
\end{theorem}

After eliminating extra variables with the proposed program transformation, we can then emulate the instantiation of variables performed by a narrowing procedure by just replacing free variables with $\mi{gen}$, thus lifting any term rewriting derivation starting from an arbitrary instance of an expression to a ground c-term.

\begin{theorem}[Lifting]\label{thGenAdeq}
For any program $\prog$, $e, e' \in \mi{Exp}$ such that $e'$ is ground:\\
\textbf{Soundness} $\gt{\prog} \vdash e[\overline{X/\gen}] \rw^* e'$ implies $\exists \sigma \in \mi{Subst}$ such that $\prog \vdash e\sigma \rw^* e''$ for some $e'' \in \mi{Exp}$ such that $|e'| \ordap |e''|$  with  $\mi{dom}(\sigma) = \overline{X}$. As a consequence, if $e' = t \in \mi{CTerm}$ then ${\prog} \vdash  e\sigma \rw^* t$.\\
%
\textbf{Completeness} For any $\sigma \in \mi{Subst}$ we have that $\prog \vdash e\sigma \rw^* e'$ implies $\gt{\prog} \vdash e[\overline{X/\gen}] \rw^* e''$ for some $e'' \in \mi{Exp}$ such that $|e'| \ordap |e''|$  with  $\overline{X} = \mi{dom}(\sigma)$. As a consequence, if $e' = t \in \mi{CTerm}$ then $\gt{\prog} \vdash  e[\overline{X/\gen}] \rw^* t$.
\end{theorem}

\section{Maude prototype}\label{maude}

We present in this section our prototype; much more information
can be found at~\url{http://gpd.sip.ucm.es/snarrowing}. 
%
%
%
%
%
%
The prototype is started by typing \texttt{loop init-s .}, that initiates an input/output loop where programs
and commands can be introduced.
These programs have syntax \texttt{smod NAME is STMNTS ends}, where \verb"NAME" is the identifier
of the program and \verb"STMNTS" is a sequence of constructor-based left-linear rewrite rules.
For instance, Example~\ref{ex:rules} would be written as
follows:

\vs
{\codesize
\begin{verbatim}
(smod CLERKS is
  branches -> madrid ? vigo .
  employees(madrid) -> e(pepe, men) .
  employees(madrid) -> e(maria, men) .
  employees(vigo) -> e(pilar, women) ? e(luis, men) .
  search(e(N,S)) -> p(N,N) .
ends)
\end{verbatim}
}
\vs

\shorttxt{\begin{figure}[thb]
\begin{center}
\framebox[.775\textwidth]{
{\codesize
\texttt{
\begin{tabular}{c|c}
\begin{tabular}{l}
(smod IPL is \\[-.25cm]
~ f(c(X),Y) -> h(X,Y) .\\[-.25cm]
~  h(X, c(Y)) -> g(X,Y) .\\[-.25cm]
~  g(0,1) -> 2 .\\[-.25cm]
ends)\\
\end{tabular}
& 
\begin{tabular}{l}
Maude> (eval-gen f(X,X) .) \\[-.25cm]
Result: 2
\end{tabular} \\
\end{tabular} 
}
}
}
\end{center}
\vspace{-.5cm}
    \caption{Using the prototype}
    \label{fig:exPrototype}
\end{figure}}



\noindent where upper-case letters are assumed to be variables. We can evaluate terms with variables
with the command \verb"eval-gen", that transforms each variable in the
term into the \verb"gen" constant described above and evaluates the thus
obtained expression in the module extended with the \verb"gen" rules:

\vs
{\codesize
\begin{verbatim}
Maude> (eval-gen search(X,X) .)
Result: p(madrid, madrid)
\end{verbatim}
}
\vs

\noindent That is, the tool first finds a result with the same value for the two elements of the pair.
We can ask the system for more solutions with the
\texttt{next} command until no more solutions are found, which will reveal pairs with different values:

{\codesize
\begin{verbatim}
Maude> (next .)
Result: p(madrid,vigo)
\end{verbatim}
}

%

Finally, the system combines the on-demand strategy with two different search strategies:
depth-first and breadth-first, and allows the user to check the trace in order to see how
the generators are instantiated. 
We will show in the following section how to use these commands.

\subsection{Looking for alternatives\label{subsec:party}}

We present here a more complex example, which introduces how to use our tool
to search for different paths leading to the solution.
This example presents a simplified version of the intruder protocol introduced
in~\cite{sNarrICTAC12}, which is also executable with the generators approach presented here
and is available at ~\url{http://gpd.sip.ucm.es/snarrowing}.

The module \verb"PARTY" below describes the specification of a party. Our goal in this party is
to have fun, so we define the function \verb"success", which receives a set of friends \verb"F"
and a set of elements that we already have. It is reduced to the function
\texttt{haveFun} applied to the set obtained after calling to our friends:

{\codesize
\begin{verbatim}
(smod PARTY is

  success(F, S) -> haveFun?(makeCalls(F, S)) .
\end{verbatim}
}

The function \texttt{haveFun} is reduced to \texttt{tt} (standing for the value \texttt{true}) when
it receives the constant \verb"fun":

{\codesize
\begin{verbatim}
haveFun(fun) -> tt .
\end{verbatim}
}

The function \verb"makeCalls" combines the current items with the ones obtained by making
further calls using the new items obtained by offering your items to your friends:

{\codesize
\begin{verbatim}
  makeCalls(F, S) -> S ? makeCalls(F, makeAnOffer(F, S)) .
\end{verbatim}
}

%

We can reach different results by using \verb"makeAnOffer". First, it is possible to
combine the current items to obtain a new one:

{\codesize
\begin{verbatim}
  makeAnOffer(F, S) -> combine(S, S) .
\end{verbatim}
}

This combination, achieved by the \verb"combine" function, generates a \verb"burger"
from \verb"bread" and \verb"meat", and \verb"fun" when a \verb"burger" and a \verb"videogame"
are found:

{\codesize
\begin{verbatim}
  combine(bread, meat) -> burger .
  combine(burger, videogames) -> fun .
\end{verbatim}
}

Another possibility is to call a friend and show him the items we have obtained so far:

{\codesize
\begin{verbatim}
  makeAnOffer(F, S) -> call(F, S) .
\end{verbatim}
}

This \verb"call" depends on the friend we call. We present below the different possibilities:

{\codesize
\begin{verbatim}
  call(enrique, drink) -> music .
  call(adri, meat) -> bread .
  call(rober, music) -> videogames .
  call(nacho, videogames) -> music .
  call(juan, food) -> drink .
ends)
\end{verbatim}
}

Once this module is loaded into the interpreter, we indicate that we want to activate the
path. In this way, we can explore the different ways to reach the values:

{\codesize
\begin{verbatim}
Maude> (path on.)
Path activated.
\end{verbatim}
}

We also set the exploration strategy to \emph{breadth first}, so the tool finds the shortest
solutions first:

{\codesize
\begin{verbatim}
Maude> (breadth-first .)
Breadth-first strategy selected.
\end{verbatim}
}

We can now look for solutions to the \verb"success" function, using a variable as argument:

{\codesize
\begin{verbatim}
Maude> (eval-gen success(F, S) .)
Result: tt
\end{verbatim}
}

We can now examine the path traversed by the tool to reach the result as follows:

{\codesize
\begin{verbatim}
Maude> (show path .)
haveFun(makeCalls(gen,gen))
--->
haveFun(gen ? makeCalls(gen,makeAnOffer(gen,gen)))
--->
haveFun(gen)
--->
haveFun(fun)
--->
tt
\end{verbatim}
}

It shows how it simply requires start with \verb"fun" to obtain \verb"fun" at the
end. Since this answer is not useful we look for the next one:

{\codesize
\begin{verbatim}
Maude> (next .)
Result: tt

Maude> (show path .)
haveFun(makeCalls(gen,gen))
--->
haveFun(gen ? makeCalls(gen,makeAnOffer(gen,gen)))
--->
haveFun(makeCalls(gen,makeAnOffer(gen,gen)))
--->
haveFun(makeAnOffer(people,gen) ?
        makeCalls(makeAnOffer(people,makeAnOffer(people,gen))))
--->
haveFun(makeAnOffer(people,gen))
--->
haveFun(combine(gen,gen))
--->
haveFun(combine(burger,gen))
--->
haveFun(combine(burger,videogames))
--->
haveFun(fun)
--->
tt
\end{verbatim}
}

In this case we
would require to start having a \verb"burger" and \verb"videogames", so they can
be combined in order to reach the \verb"fun". In this case no friends were required.
However, the next search (where we just show the last steps) requires a \verb"burger",
\verb"music", and our friend \verb"rober":

{\codesize
\begin{verbatim}
...
haveFun(combine(gen,call(gen,gen)))
--->
haveFun(combine(burger,call(gen,gen)))
--->
haveFun(combine(burger,call(rober,gen)))
--->
haveFun(combine(burger,call(rober,music)))
--->
haveFun(combine(burger,videogames))
--->
haveFun(fun)
--->
tt
\end{verbatim}
}

We can keep looking for more results
until we find the one we are looking for or we reach the limit on the number of
steps (which can be modified by means of the \verb"depth" command).

\subsection{Implementation notes}

We have implemented our prototype in Maude~\cite{maude-book},
a high-level language and high-performance
system supporting both equational and rewriting logic computation
for a wide range of applications. Maude modules correspond
to specifications in \emph{rewriting logic}~\cite{Meseguer92-tcs}, a simple
and expressive logic which allows the representation of many models
of concurrent and distributed systems. This logic is an extension of
equational logic; in particular, Maude \emph{functional modules} correspond to
specifications in \emph{membership equational logic}~\cite{BouhoulaJouannaudMeseguer00},
which, in addition to equations,
allows the statement of \emph{membership axioms} characterizing the elements
of a sort.
Rewriting logic extends membership equational logic by adding rewrite
rules, that represent transitions in a concurrent system.
This logic is a good semantic framework for formally specifying
programming languages as rewrite theories~\cite{semanticsProject};
since Maude specifications are executable, we obtain an interpreter for the language
being specified.

Exploiting the fact that rewriting logic is reflective~\cite{ClavelMeseguerPalomino07},
an important feature of Maude
is its systematic and efficient use of reflection
through its predefined \texttt{META-LEVEL} module~\cite[Chapter~14]{maude-book},
a characteristic that allows many advanced metaprogramming and metalanguage applications.
This powerful feature allows access to metalevel entities such as
specifications or computations as usual data. In this way, we define the syntax of the modules
introduced by the user, manipulate them, direct the evaluation of the terms
(by using the on-demand strategy natural narrowing~\cite{escobar04}), and implement
the input/output interactions in Maude itself.


\section{Concluding remarks and ongoing work}\label{sect:conclusions}

In this work we have proposed and formally proved the adequacy of a technique for lifting
term rewriting derivations from an arbitrary instance of an expression to a constructed
term---or the outer constructed part of any expression---using left-linear constructor
systems. It is
based on the generator technique from the field of functional-logic
programming~\cite{DiosLopez07,AntoyH06Extra}, but adapted to the different
semantic context of term rewriting~\cite{rodH08}.
For proving the adequacy of the proposed technique we have employed the semantics for
constructor systems defined in~\cite{Lopez-FraguasRS09-RTA09} as the main technical tool.
This way we have put
the semantics in practice by using it for solving a technical problem that was not stated
in the original paper. Along the way we have extended the semantics to support extra variables
in rewriting rules, as those are very frequent when using narrowing, which is the context of
the present paper. To do that we have made the necessary adjustments to the formulation
of the semantics and to the proofs for its properties.

%
A \longtxt{fundamental} limitation of the generators is that they can only be used for reaching ground c-terms \longtxt{or the outer constructed part of expressions}. \longtxt{This limitation can be \longtxt{somewhat} mitigated by reducing the reachability to a non-ground value to the reachability of a ground value: for example to test for $e \rw^* c(X)$ we can define a new function $f$ by the rule $f(c(X)) \tor \mi{true}$ and then test for $f(e) \rw^* \mi{true}$. Anyway this is a partial solution, and moreover the instantiation of free variables corresponding to the evaluation of $gen$ cannot be obtained by a transformation in that line, for example by evaluating $(f(X), X)$ in the previous example, due to the aforementioned loss of sharing between different occurrences of the same variable. This latter limitation could only be possibly overcomed by using some metaprogramming capabilities of the rewriting engine used to implement this technique.
The generators technique has been used in practical systems, for example as the basis for an implementation of  the functional-programming language Curry~\cite{BrasselHPR11KicsDos}. There the information provided by a Damas-Milner like type system is used to improve the efficiency, because instead of just one universal generator, like in our proposal, several generators are used, one for each type, which results in a great shrink of the search space for the evaluation of generators. One could argue that our generators are fundamentally equivalent to defining a generator $genE$ that could be reduced to any expression, and then replacing each free or extra variable with $genE$, which would be trivially complete. Nevertheless, in our approach the search space for generators is significantly smaller, especially when combined with type information.}
\shorttxt{We provide an efficient implementation in Maude
using the on-demand strategy natural rewriting~\cite{escobar04}, which allows us
to study their expressivity and possible applications.
Along the way we have extended the semantics for constructor systems from~\cite{Lopez-FraguasRS09-RTA09} to support extra variables in rewriting rules, as those are very frequent when using narrowing. In the future we plan to improve our implementation by using the reflection capabilities of Maude to collect the evaluation of generators, so we would be able to present a computed answer for generators derivations, and use type information to improve the performance. }

\longtxt{The system has been implemented in a Maude prototype that allows us to study their expressivity
and possible applications. This prototype uses the on-demand strategy natural rewriting~\cite{escobar04}, thus providing an efficient implementation.}

Regarding future work, we plan to improve our implementation by using the reflection capabilities of Maude to collect the evaluation of generators, in order to be able to present a computed answer for generators derivations, instead of relying on the trace to extract this information.

\bibliographystyle{eptcs}
\bibliography{references}

\end{document}